# SFAMSS: A Secure Framework For ATM Machines Via Secret Sharing

Zeinab Ghafari[1], Taha Arian[2] and Morteza Analoui[3]

[1] School of Computer Engineering Iran University of Science and Technology, Tehran,Iran
z_ghafari@cmps2.iust.ac.ir

[2] School of Computer Engineering Iran University of Science and Technology, Tehran,Iran
t_arian@ce.sharif.edu

[3] Associate Professor of Computer Engineering Iran University of Science and Technology,Tehran,Iran
analoui@iust.ac.ir

## ABSTRACT

As ATM applications deploy for a banking system, the need to secure communications will become critical. However, multicast protocols do not fit the point-to-point model of most network security protocols which were designed with unicast communications in mind. In recent years, we have seen the emergence and the growing of ATMs (Automatic Teller Machines) in banking systems. Many banks are extending their activity and increasing transactions by using ATMs. ATM will allow them to reach more customers in a cost effective way and to make their transactions fast and efficient. However, communicating in the network must satisfy integrity, privacy, confidentiality, authentication and non-repudiation. Many frameworks have been implemented to provide security in communication and transactions. In this paper, we analyze ATM communication protocol and propose a novel framework for ATM systems that allows entities communicate in a secure way without using a lot of storage. We describe the architecture and operation of SFAMSS in detail. Our framework is implemented with Java and the software architecture, and its components are studied in detailed.

## 1. INTRODUCTION

Security protocols are applied on untrusted network to enhance their safety. ATM networks are one of the typical networks that need a high level of security to prevent the attacker doing malicious activity. ATM communication consists of several phases, such as authentication and authorization. So, establishing a comprehensive security in ATM infrastructure needs a lot of concerns. Lack of security even in one of the phases can lead to massive security breach. In order to solve this problem, we propose a new framework that includes several entities same as ATM, customer, and bank. To achieve security in the ATM networks each entity should consider the security as an important factor.

There are some studies on designing a new protocol and their attacks on ATM. In our previous paper [1] we presented two security protocols for ATM communication. In this paper, our motivation is to introduce a new framework that includes registration, authentication, and authorization. Firstly, the user and ATM register for bank's services. The bank generates authentication information and distributes it among them. Bank also assigns some privileges to users, such as the amount of money he could transfer by means of ATM. Secondly the bank authenticates user by ATM, and finally the bank authorizes user's request for a service, based on the user's privileges. It is possible that a user be authenticated by the bank; however, he is not authorized for a particular service. We implement our framework by using Java programming language on the Eclipse platform. Our framework will investigate as a secure framework for ATM machines via secret sharing (SFAMSS) in the rest of paper.

Secret sharing schemes (SSS) are perfect for storing information that is highly confidential and critical. For instance encryption keys, missile launch codes, and bank account information are critical information that should maintain in a secure way. This information must be protected highly confidential, as their exposure could be harmful. Traditional methods for encryption are not suited for achieving high levels of confidentiality. Secret sharing aims at efficiently sharing a secret among a number of entities, and the secret can be recovered by entities shares. Hence, there are several applications of secret sharing in computer science such as [2],[3] in cloud and [4],[5] in data outsourcing.

The rest of the paper is organized as follows: Section 2 analyzes related work and Section 3 describes the framework in particular, highlighting its main architecture and protocol details. Then, Section 4 illustrates the implementation, and a case study in detail. Section 5 reports the protocol analysis. Finally, Section 6 concludes the paper.

## 2. RELATED WORKS

There is a considerable amount of literature on designing a new secure framework in network communication. Few studies have been published on banking systems. In [6] the authors proposed a framework based on a smart card that allows entities to realize secure transactions. The proposed solution uses smart cards to store keys and perform cryptographic algorithms in e-business transactions. Robinson P. et.al. [7] reported a new framework for a secure protocol. The paper describes the implementation of a deterministic and fair non-repudiation protocol. Malin and Sweeney [8] presented the STRANON protocol that enables locations to cooperate such that health records are not revealed. They developed a general high-level framework for STRANON. Authors in [9] presented a new framework intended to extract FSA(Finite State Automaton) specifications of network protocol implementations and test it for implementation flaws. They constructed the framework using Java. Mittra [10] developed a novel framework for scalable secure multicasting this protocol can be used to achieve a variety of security objectives in communications. In [11] the authors investigated a new framework for network management protocol. They proposed an architecture composed of mobile agents. Camenisch et.al[12] proposed a framework for cloud security storage. They applied secret sharing scheme to prevent unauthorized access to cloud data. They provide a secure analysis for their protocol by using zero-knowledge proof.

## 3. SFAMSS FRAMEWORK

In this paper, we implement the authentication and secure communication framework for a distributed ATM systems, consisting of a bank server, a number of ATMs and users. Since the computer system at the bank is connected through an insecure communication channels to the ATM, applying secure mechanisms in the communication is essential. The insecure communication channels are subjected to attacks by active and passive malicious attacker. Messages may not be removed by an attacker. However, confidentiality of messages might be breached, and new message might be generated by the attacker. Thus, the authorship and content of messages that transit in the insecure communication link should be considered suspect. Customers use ATMs to make queries such as withdrawals and balance inquiries involving their accounts. Attackers must be prevented from interfering with these actions.

Communications between user, ATM and bank consists of the following steps:
1. The user inserts his smart card into the ATM.
2. The ATM informed the customer for a password.
3. The user selects a request to be completed by ATM, and that request is performed by the bank and the result is shown by the ATM to the user.

### 3.1. SFAMSS Architecture

In this section, we considered that three parties in ATM system do not trust each other among interaction network. So, to protect interacts among parties, communication should be confidential and repudiable. Confidentiality is the property that a message cannot be accessed by unauthorized entities. Non-repudiation is the property that an entity cannot repudiate the message that he send before. In this section, we present a new protocol that guarantees non-repudiation and confidentiality. Figure 1 illustrates SFAMSS architecture with SSS component that changes the authentication process of ATM systems. Using SSS helps entities to authenticate each other without trusted third party existence.

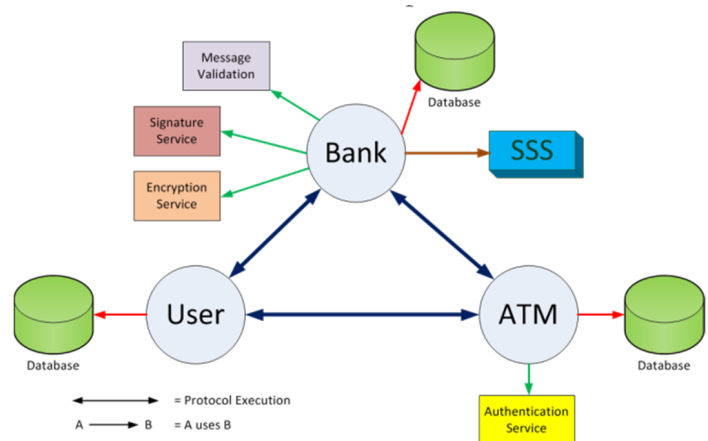

Figure 1-SFAMSS architecture

### 3.2. Protocol details

We have a certificate authority that issues certificates to all entities, so, before the registration phase all of the entities have their certificate and private key. In all of the protocol descriptions, $E(key_{AB})(msg)$ means message (msg) is encrypted by key ($key_{AB}$). $E_{pu}$, $E_{pr}$ are referred to entity E public key and private key. $K_S$ is a session key which is established between user and bank. The ATM communicates with the bank by means of a protocol that meets the following requirements:
1. ATM authenticates the user.
2. ATM sends authentication information to the bank.
3. It preserves the integrity and confidentiality of communications between the bank and ATM.
Messages that are sent by the ATM to the bank provide evidence that every ATM-initiated action was, in fact, initiated by the user. The bank server responses to user by ATM. It also records information in an audit log for later use in justifying its past actions to the user. The audit log that is stored in bank servers is vulnerable to network attacks. So data written to the log should be encrypted to prevent a confidentiality breach. In addition to log, bank should keep accurate information about users and ATMs secretly. Since entities sign their messages, they cannot repudiate it.

$bank_{pb}$: Banks public key
$bank_{pr}$: Banks private key
$K(u_1; bank)$: Session key between bank and $u_1$
$u_{1pb}$ : Users public key
$u_{1pr}$ : Users private key The private key of the user is stored securely in the credit card
$ATM_{pb}$: ATMs public key
$ATM_{pr}$: ATMs private key. The private key of the ATM is stored securely at the ATM.

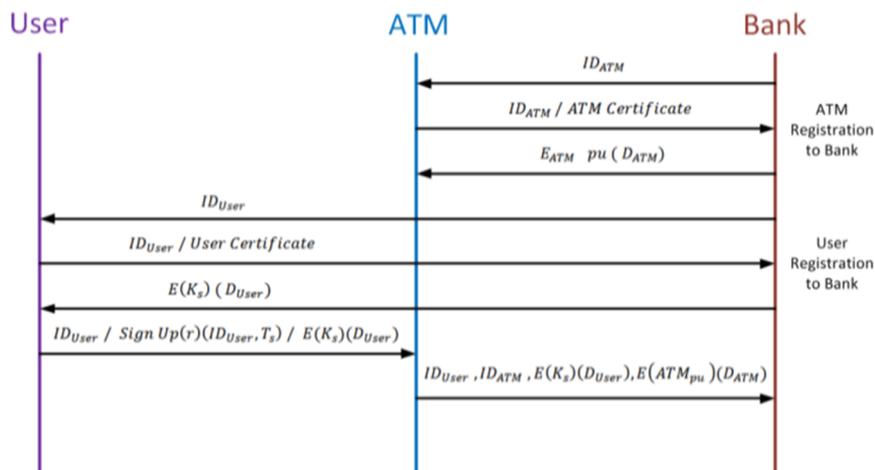

Figure 2-Message Exchange in the selected scenario

Our scenario is illustrated in Figure 2 and we will investigate it step by step.

### 3.2.1. Registration of the ATM with the bank

$BANK \rightarrow ATM : ID_{ATM}$

$ATM \rightarrow BANK: ID_{ATM}, ATM_{certificate}$

[BANK generate function $F(x)$ & calculate $D_{ATM} = (ID_{ATM}, F(ID_{ATM}))$ ]

$BANK \rightarrow ATM : E(ATM_{PU})(D_{ATM})$

### 3.2.2. Registration of the User with the bank

$BANK \rightarrow USER : ID_{USER}$

$USER \rightarrow BANK : ID_{USER}, USER_{Certificate}$

[ BANK generate random $r_{USER}$ and store $D_{BANK,USER} = (0, r_{USER})$ &
calculate $f_{new}(x) = f(x) + r_{USER}$
& $D_{USER} = (ID_{USER}, f_{new}(ID_{USER}))$ ]

$BANK \rightarrow USER : E(K_S)(D_{USER})$

### 3.2.3. Authentication of User to ATM

$$\text{USER} \rightarrow \text{ATM}: \text{ID}_{\text{USER}}, \text{SIGN}(\text{USER pr})(\text{ID}_{\text{USER}}, T_S), E(K_S)(D_{\text{USER}})$$

$$\left[\begin{array}{l}\text{ATM requests user certificate from BANK and}\\\quad\text{after that ATM verify the signiture}\\\text{by the user bublic then ATM authenticate USER}\end{array}\right.$$

### 3.2.4. Authentication of ATM and User to Bank

$$\text{ATM} \rightarrow \text{BANK}: \text{ID}_{\text{USER}}, \text{ID}_{\text{ATM}}, E(K_S)(D_{\text{USER}}), E(\text{ATM}_{\text{PU}})(D_{\text{ATM}})$$

Bank decrypts $D_{\text{USER}}$ with session key between Bank and User and decrypts $D_{\text{ATM}}$ with $\text{ATM}_{\text{PU}}$

and calculate $D'_{\text{ATM}} = (\text{ID}_{\text{ATM}}, F(\text{ID}_{\text{ATM}}) + r_{\text{USER}})$
generate $F'_{\text{NEW}}(x)$ with $D_{\text{USER}}, D'_{\text{ATM}}, D_{\text{BANK}}$

$$F'_{\text{NEW}}(x) = ? \ F_{\text{NEW}}(x)$$

## 4. Implementation

The UML design goal is to represent the several components and technologies that will be used to build a framework. We implemented protocol role with Java language, in the eclipse platform. The UML diagram are illustrated in this hyperlink.

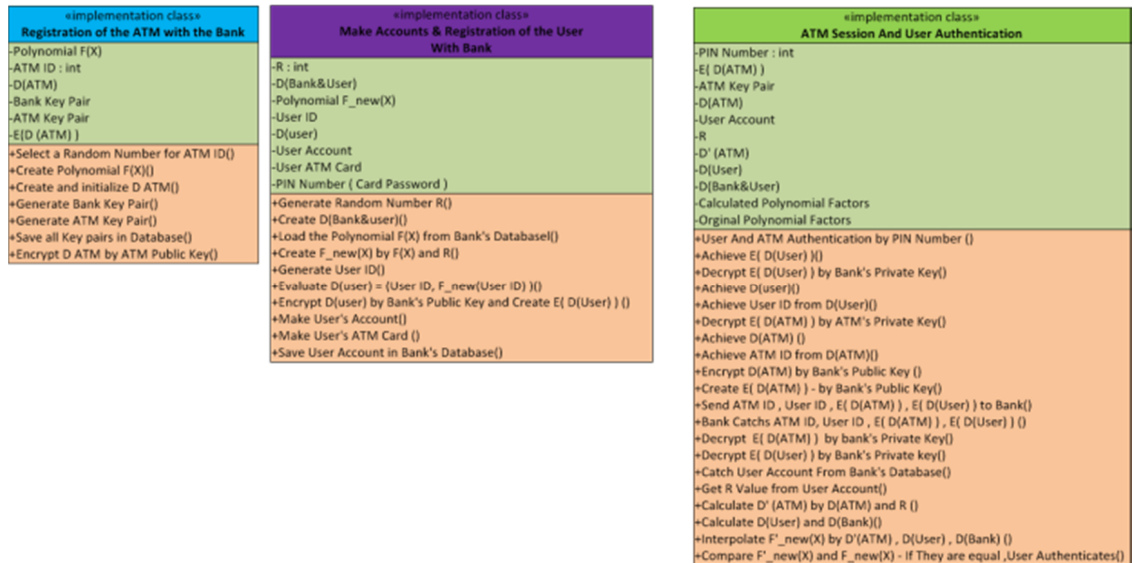

Figure 3- Implementation classes

### 4.1. Case study

In the registration phase of ATM and the bank, first bank generates a key pair for ATM based on the RSA algorithm. Second, he stores them in the bank database. Bank generates polynomial function F(x) then creates D(ATM) with it. Bank encrypts D(ATM) with ATM public key and sends it to ATM.

| Algorithm1 :registration ATM with BANK |
|---|
| ATM establishes communication with Bank |

```
SET bank_pu, bank_pr
SET F(X) = x^a + bx + c
READ ID_ATM
SEND ATM-certificate to BANK for verification
if failure then Display("Bad certificate") and exit
end if
SET D_ATM=F(ID_ATM)
SEND E (D_ATM) to ATM
```

In phases of user registration with the bank, at the first we choose a random number R for each user then generate D(bank,user) that is equal to R. Now, calculate $F_{new}(x)$ which $F_{new}(0)=R$ and choose a random number I for ID user. Bank generates D(user)= $F_{new}(i)$ then encrypt and puts it on a smart card.

| Algorithm2 :registration USER with BANK |
|---|
| SET user_pu, user_pr |
| SET random number (R) |
| SET D(BANK,USER)=(0,R) |
| SET Fnew(X)=F(X)+R |
| SET random number (I) |
| SET Duser=Fnew(I) |
| SEND E(Duser) |

In the phases of user to bank authentication, first user authenticates with ATM. User sends password number to ATM with his smartcard. ATM checks if the password is correct. ATM extracts the Duser and Datm.

| Algorithm3 authentication between user and ATM |
|---|
| Display("Please enter password") |
| passwd= get input from user() |
| verifyUsingSC(passwd) |
| if verified then Authentication Successful |
| READ Duser,Datm |
| else Authentication Failed |

In the user and ATM authentication phase with the bank, at the first, bank retrieves D(bank,user) whit user ID. Then, bank calculates D'(ATM). Finally, bank interpolates new function F'new(x) with three points D'(ATM), D(user), D(bank,user). To verify the authentication among entities bank checks if $F'_{new}(x) == F_{new}(x)$.

| Algorithm4 authentication among BANK,ATM,USER |
|---|
| Bank read D(bank,user) |
| SET D'atm=Datm+R |
| SET interpolation F'new with(D'atm,Duser,D(bank,user)) |
| IF F'new(x)=Fnew(x) then authentication successful |
| else |
| Authentication failed |

# 5. Protocol Security Analysis

Security properties required by ATM systems include confidentiality, authentication, integrity and non-repudiation.

**Confidentiality**

Confidentially means guaranteeing data and important user information not to be accessed by unauthorized users and aliens that usually is performed using cryptography techniques. We encrypt messages such as shares to establish confidentiality in the communication. Thus, dues to the shares that we have considered exclusively for each entity, to access the main secret data, it is required that all the entities must be present, which results in establishing the confidentiality. Also the bank and ATM key pair are used to encrypt messages. This key pair used between the bank and the user prevents the ATM from accessing the information shared between these two entities, which in turn results in the confidentiality. Confidentiality of our algorithm depends on the number of key bits and encryption algorithm. We use RSA as a robust encryption algorithm. There is a trade-off between number of key bits and performance of the algorithm. If the number of key bits is large, protocol security is higher but encryption and decryption process may take longer.

**Integrity**

Integrity is needed to prevent and discover redundancy, modification, and deletion of data. While registering an ATM and a user in a bank, two independent certificates are issued for these two entities. Both certificates are signed by the user. To establish integrity, we used digital signature in the protocol communication. In SFAMSS, if an attacker tries to change part of the message, receiver can detect the changes by verifying the signature of the message.

**Non-Repudiation**

Non-repudiation prevents the sender from denying the transmission of her message. We create this mechanism using the signature in the scenario. Since all the signatures in our protocol use private key of the sender, nobody can repudiates the message that he sents before. Because the only entity who has the private key is the sender of a message.

**Freshness**

In our protocol, we use the timestamp to keep our message freshness. It ensures that the messages are recent, and it ensures that no attacker replayed old messages. Hence, if an attacker uses her previous messages to access the information, the bank detects this situation, due to the existence of the timestamp.

# 6. Conclusion

Our contribution in this paper is to present a framework for ATM systems. Our framework ensures confidentiality and non-repudiation and integrity in communication between bank, ATM, and user. In the registration phase of ATM, bank, and user, secret shares are distributed among users by bank SSS module. In the authentication phase of ATM, bank, and user, shares are aggregated to generate a polynomial function. Bank checks the polynomial function to verify if entities are authenticated. Since we use digital signature and encryption algorithms to enhance the security of ATM communication protocol. We conclude that secret sharing schemes are suitable to use in ATM systems. In the future work, we will analyse the protocol with protocol verifier tools such as proverif [13].